\documentclass[showpacs, showkeys,twocolumn,preprintnumbers,floatfix,amsmath,amssymb]{revtex4}
\usepackage{natbib}
\usepackage{amsmath,amsthm}
\usepackage{float}
\usepackage{graphicx}
\usepackage{caption}
\usepackage{subfigure}
\usepackage{tabularx}
\usepackage{threeparttable}
\usepackage{dcolumn}
\usepackage{bm}
\usepackage{color,epsfig,multirow}
\usepackage{array}
\usepackage{hyperref}
\usepackage{algorithm}
\usepackage{algorithmic}
\usepackage{booktabs}
\usepackage{threeparttable}

\numberwithin{equation}{section}

\DeclareMathOperator{\var}{var}

\newtheorem{theorem}{Theorem}[section]

\theoremstyle{definition}

\theoremstyle{remark}
\newtheorem{remark}[theorem]{Remark}

\allowdisplaybreaks[4]

\begin{document}
\preprint{Preprint}

\title{Random-batch list algorithm for short-range molecular dynamics simulations}
\author{Jiuyang Liang$^1$}\thanks{liangjiuyang@sjtu.edu.cn}
\author{Zhenli Xu$^{1,2}$}\thanks{xuzl@sjtu.edu.cn}
\author{Yue Zhao$^1$}\thanks{sjtu-15-zy@sjtu.edu.cn}
\affiliation{$^1$School of Mathematical Sciences, Shanghai Jiao Tong University, Shanghai 200240, China\\
$^2$Institute of Natural Sciences and MoE Key Lab of Scientific and Engineering Computing, Shanghai Jiao Tong University, Shanghai 200240, China}
\date{\today}
\begin{abstract}
We propose a fast method for the calculation of short-range interactions in molecular dynamics
simulations. The so-called random-batch list method is a stochastic version of the classical neighbor-list method to avoid the construction of a full Verlet list,
which introduces two-level neighbor lists for each particle such that the neighboring particles are located in core and  shell regions, respectively.
 Direct interactions are performed in the core region. For the shell zone,
we employ a random batch of interacting particles to reduce the number of interaction pairs.
The error estimate of the algorithm is provided. We investigate the Lennard-Jones fluid by molecular dynamics simulations, and show that this novel method can significantly
accelerate the simulations with a factor of several fold without loss of the accuracy.
This method is simple to implement, can be well combined with any linked cell methods to further speed up and scale up the simulation systems, and can be straightforwardly extended to other interactions such as Ewald short-range part, and thus it is promising for large-scale molecular dynamics simulations.
\end{abstract}



\maketitle

\section{Introduction}

Molecular dynamics (MD) has become one of the most popular simulation tools in studies of systems at the nano/micro scale such as chemical physics, biological and soft matter systems.
The MD measures equilibrium and dynamical properties of physical systems by ensemble average of particle configurations \cite{lifshitz2013statistical,allen1987,frenkel2002book}
produced by integrating the Newton's equations of motion.
This algorithm is simple and highly parallel, whereas the calculation of the nonbonded forces on particles
is the computational bottleneck and remains wide interest in algorithm development and optimization.
These nonbonded interactions include the Lennard-Jones (LJ) and Coulomb potentials. Coulomb interactions are usually calculated by the Ewald-type lattice summation
where the smooth part is treated on uniform mesh via fast Fourier transform \cite{darden1993particle,Philippe2000} and the real part decays rapidly, i.e., becomes short-range interactions,
similar to the LJ interaction. With the increasing optimization of algorithms for long-range forces \cite{JinLiXuZhao2020}, the proportion of the CPU cost
on short-range interactions tends to play essential contribution, and thus it will be significant to develop novel algorithms for accelerating short-range calculations.

Fast techniques for short-range interactions have been extensively discussed in literature \cite{verlet1967computer,hockney1974quiet,brown2011implementing,pall2013flexible,plimpton1995fast,howard2016efficient,yao2004improved, heinz2004fast, meloni2007efficient, welling2011efficiency,gonnet2007simple,kadau2006molecular,neumann2019petaflop,NikolaTchipev2019,Yang16JCC}.
A simple idea to reduce the computational cost is the cutoff scheme which truncates the interaction potential between a pair
of particles at a radial cutoff distance and ignores the pairs of larger distances. A neighbor list which stores neighboring particles can be efficiently built by the cell list such that the linear scaling of the computational complexity is achieved.
The Verlet list \cite{verlet1967computer} introduces an additional larger cutoff radius to reduce the frequency of neighbor list establishment. The prefactor of the linear-scaling neighbor list algorithm depends on the average number of particles within the cutoff radius, which can be large for heterogeneous systems due to a big radius having to be used.
Scalable algorithms that work well on modern computer architecture have been frequently reported recently. Many of them devote to the improvement of the linked cell list (LCL) algorithm employed for the Verlet list construction \cite{yao2004improved, heinz2004fast, meloni2007efficient, welling2011efficiency,gonnet2007simple,kadau2006molecular,neumann2019petaflop,NikolaTchipev2019}, attempting to the reduction of the memory access and computation cost. Szil{\'a}rd {\it et al.} \cite{pall2013flexible} developed
an algorithm for SIMD parallelization based on grouping a fixed number of particles into spatial clusters, which significantly reduces the costly shuffling
operations \cite{pennycook2013exploring}. The requirements of high level of parallelism and sensibility of memory access patterns on GPU MIMD architecture also attract
much interest \cite{howard2016efficient,proctor2013gpu,Yang16JCC}. Most of these algorithms are implemented into broadly-used MD packages, such as GROMACS \cite{pronk2013gromacs}
and LAMMPS \cite{plimpton2012computational,plimpton1995fast}.

In this work, we report a novel random-batch list (RBL) method for the calculation of short-range interactions in MD simulations.
The RBL introduces two neighbor lists in domains of core-shell structure around
each particle. The interactions with the core particles are calculated directly. For particles in the shell,
the random-batch idea \cite{jin2020random,li2020random} is introduced to build a minibatch of particles randomly taken from the shell zone via stenciled cells.
The central particle only interacts with these particles with scaled interacting forces. Theoretically, our algorithm provides an unbiased
estimate of the force acting on each particle. Our calculation on LJ fluids shows that the core radius of the RBL can be much smaller than the cutoff radius
in the classical scheme which combines the Verlet and cell lists.
Since the size of the batch can be much smaller than the number of particles in the shell zone,
the algorithm significantly reduces the computational complexity as well as the storage reduction with several folds.

The rest of the paper is organized as follows. In Section 2, we introduce the LJ model, and then describes the RBL in details.
In Section 3, we perform numerical calculations for benchmark problems of LJ fluids with different temperatures and
densities, and validate the superior performance the method. Conclusions are given in Section 4.

\section{Method}\label{Sec::Med}
We consider a system composed of $N$ particles located at $\bm{r}_i$, $i=1,\dots, N,$  within a cubic box of side length $L$.
One of the most widely used intermolecular potentials in classical many-body simulations is the so-called LJ $12-6$ potential,
\begin{equation}\label{12-6}
U_{ij}=4\varepsilon \left[\left(\frac{\sigma}{r_{ij}}\right)^{12}-\left(\frac{\sigma}{r_{ij}}\right)^6\right],
\end{equation}
where $\varepsilon$, $\sigma$, and $r_{ij}$ denote the depth of the attractive well, the interparticle distance at which the potential changes sign, and the Euclidean distance (or the minimum-image distance \cite{frenkel2002book} if the system is periodic) between particle $i$ and particle $j$, respectively. 
In practical calculation, a cutoff radius $r_\mathrm{s}$ is introduced such that pair contributions beyond this distance are ignored.
An energy shift is often added to maintain the continuity of potential energy such that the energy becomes zero at the distance $r_\mathrm{s}$.
When the temperature is high and the density is low, a small cutoff radius can be used, e.g., $r_\mathrm{s}=2.5\sigma$
can often predict very nice results \cite{frenkel2002book}.
But at the gas-liquid coexistence state, as in the LJ fluid model \cite{johnson1993lennard}, one needs to choose large $r_\mathrm{s}$ to correctly describe the liquid phase.
The particle number in the neighbor list will be very large for such heterogeneous systems to achieve high accuracy, limiting the computational efficiency for large scale problems.

The RBL method proposed in this paper constructs the neighbor list which significantly reduces the number of interacting pairs.
The idea depends on the random batch strategy \cite{jin2020random} to solve stochastic differential equations
for interacting particle systems, originally from stochastic gradient descent method \cite{bottou1998online,robbins1951stochastic}. The random batch idea has been successfully used in many fields such as efficient samplings \cite{ye2021efficient,li2020stochastic}, Monte Carlo  \cite{li2020random,CiCP-28-1907,li2021splitting}, and Coulomb systems \cite{JinLiXuZhao2020,liang2021superscalable}. The random batch method approximates the force acting on each particle by the forces (with scaled strengths) due to the particles in the chosen batch, which is shown to be an unbiased estimate of the force.
For LJ systems, the issue of kernel singularity leads to uncontrolled variance. This can be treated by acceptance-rejection rule in the classical Metropolis manner
to obtain a Monte Carlo scheme \cite{li2020random}.
In the MD method, we split the domain surrounding each particle into a core region and a shell zone and employing the random batch only on the shell zone.
This strategy naturally prevents unphysical overlap of particles. The force variance is proven to be bounded and the computation cost can be dramatically reduced.
Moreover, an appropriate thermostat can be employed to further control the force variance. For example,  with the friction and diffusion terms included in Newton's equations of motion, the Langevin dynamics is described by,
\begin{equation}
	\begin{split}\label{Langevin}
		&d\bm{r}_i=\bm{v}_idt,\\
		&m_id\bm{v}_i=[-\nabla_{\bm{r}_i}U-\gamma \bm{v}_i]dt+\sqrt{2\gamma k_B T}d\bm{W}_i,
	\end{split}
\end{equation}
where $\bm{v}_i$ is the velocity of particle $i$, $\gamma$ is the reciprocal characteristic time associated to the Langevin thermostat, $k_B T$ is the thermal energy and $\{\bm{W}_i\}_{i=1}^N$ are i.i.d. Wiener processes.

The RBL introduces two cutoff radii $r_\mathrm{c}$ and $r_\mathrm{s}$ such that core-shell structured neighbor lists are constructed around each particle.
Let $r_\mathrm{c}<r_\mathrm{s}$ and $r_\mathrm{s}$ be the typical radius in the traditional cutoff scheme. We treat the core particles by direct summation and the shell particles by the random batch.
This strategy is expected to reduce the number of interacting pairs as well as remain the accuracy by resolving the kernel singularity issue.

\begin{figure}[htbp]	
	\centering
	\includegraphics[width=0.5\textwidth]{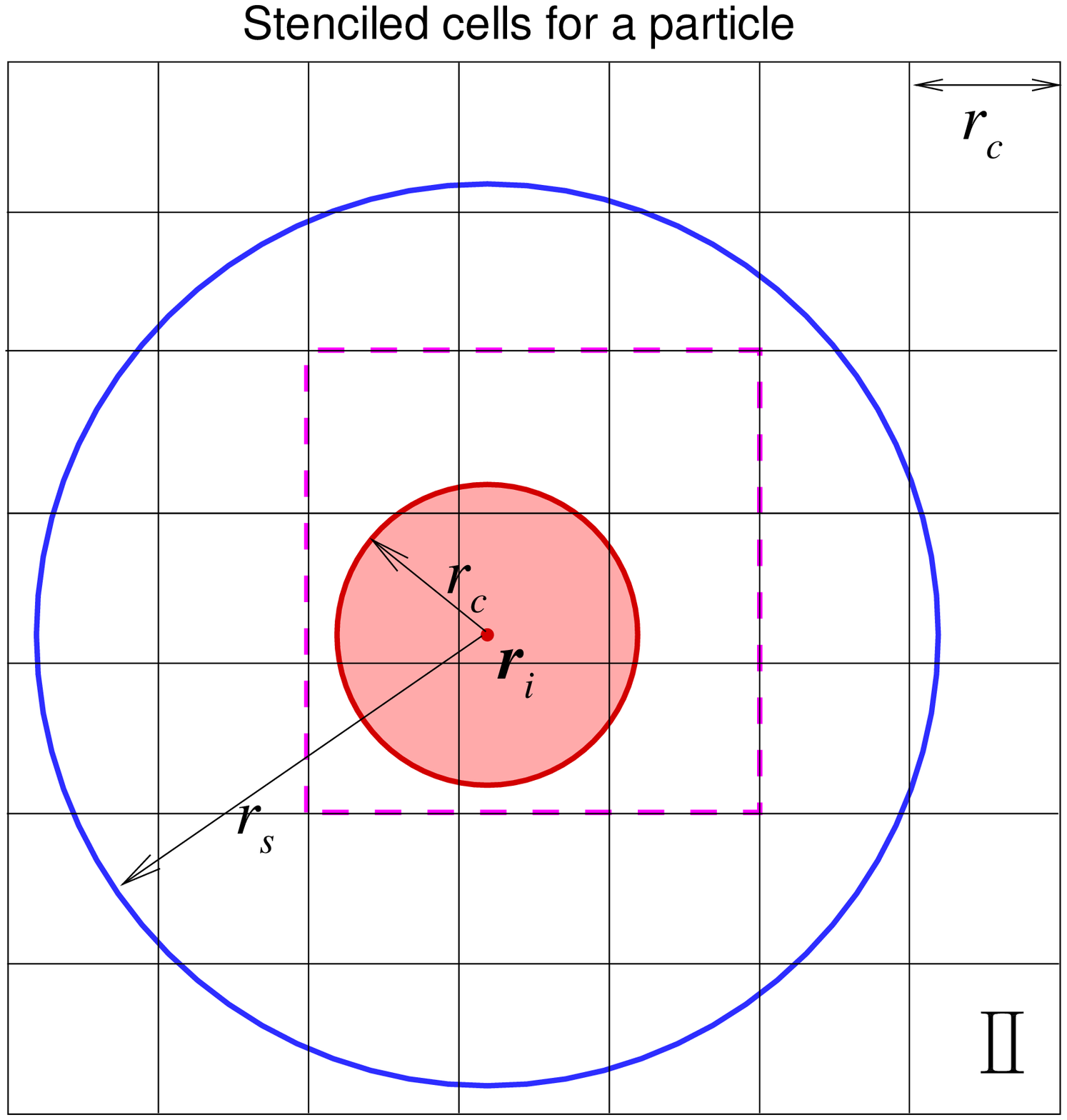}
	\caption{Schematic of the construction of the neighbor lists by stenciled cells. The core neighbor list is built by the cell lists within the dashed line (27 cells in 3D).
The random batch of particles from the shell zone is constructed by all the cells in the stencil. }
	\label{fig:neilist}
\end{figure}

Technically, the neighbor lists are obtained from the stenciled cell list. The simulation box is divided into uniform cells of edge $r_\mathrm{c}$, and the particle list in each cell is built (see Fig. \ref{fig:neilist}).
For a particle $i$, the core neighbor list is then constructed from the nearest $27$ cells by taking particles with distance less than $r_\mathrm{c}$.
For the shell particles, a stencil of cells, a combination of all neighboring particles of $i$ into uniformly sized cells of width $r_{\mathrm{c}}$), $\mathbb{I}$, is employed. The stencil is a larger box, including $(2\lfloor r_\mathrm{s}/r_\mathrm{c} \rfloor+3)^3$ cells where particle $i$ is located at the central cell.
Here, we only need to know the number of particles in the stenciled cells, $N_{\mathbb{I}}$, and it can be calculated in the step of constructing cell lists.
In practice, the edge of cell can be slightly larger than $r_\mathrm{c}$, so that updating the neighbor list is not required at every step \cite{plimpton1995fast}. If the system has large disparity in interaction lengths, the standard list is extended so that the cell width is based on the smallest cutoff radius and each particle type searches a different ``stencil'' of adjacent cells based on the largest
cutoff radius \cite{in2008accurate,howard2016efficient}.
Distances to each cell in the stencil are precomputed and the particle distance check can be skipped for many of the searched particles.

With these preparation, we discuss the force calculation on particle $i$, which is expressed as the summation of two contributions from the core and the shell
particles, $\widetilde{\bm{F}}_i=\bm{F}_{i,\mathrm{c}}+\widetilde{\bm{F}}_{i,\mathrm{s}}$. The contribution due to particles in the core list is calculated directly,
\begin{equation}\label{coreF}
\bm{F}_{i,\mathrm{c}}=\sum\limits_{r_{ij}<r_\mathrm{c}}\bm{f}_{ij},
\end{equation}
where $\bm{f}_{ij}$ is the force on particle $i$ due to particle $j$.
The force due to particles in the spherical shell zone $\widetilde{\bm{F}}_{i,\mathrm{s}}$ is calculated by using the random batch. Let $p$ be the batch size, i.e.,
we attempt to randomly choose $p$ particles from, $\mathbb{I}$, the stenciled cells. If $N_{\mathbb{I}} \leq p$, we can calculate $\widetilde{\bm{F}}_{i,\mathrm{s}}$
by taking contribution of all particles in the shell zone, and the method reduces to the classical cutoff scheme for the LJ system.
Otherwise, the $p$ chosen particles are grouped into the batch $B(i)$, and the force due to the contribution from the shell zone is approximated by
\begin{equation}\label{shellF}
\widetilde{\bm{F}}_{i,\mathrm{s}}=\frac{N_{\mathbb{I}}}{p}\sum\limits_{j\in B(i)} \widetilde{\bm{f}}_{ij},
\end{equation}
where $\widetilde{\bm{f}}_{ij}=\bm{f}_{ij}$ if particle $j$ is in the shell zone $r_\mathrm{c}<r_{ij}<r_\mathrm{s}$, and zero otherwise.
Finally, the net force of all particles is calculated and we subtract the average net force on each particle to obtain zero net force which ensures the conservation of total momentum. Note this subtraction factor is a random variable with zero expectation and bounded variance (see Appendix \ref{app:forcevar}).

Let $\bm{\chi}_{i}=\bm{F}_{i}-\widetilde{\bm{F}}_{i}$.
It is proved in Appendix \ref{app:forcevar} that the expectation of $\bm{\chi}_{i}$ is zero and the variance of $\bm{\chi}_{i}$ is bounded, and therefore our approximation is unbiased.
The core list ensures that the particles do not get too close to overlap, and the shell list is used to maintain long-range correlation between particles.
The difference between the random-batch force and the true force is small because of the accurate calculation of force in the core list.

Let us define  the Wasserstein-2 distance \cite{santambrogio2015} as
\begin{gather}\label{eq:W2}
	W_2(\mu, \nu)=\left(\inf_{\gamma \in \Pi(\mu,\nu)}\int_{\mathbb{R}^3\times\mathbb{R}^3}|\bm{x}-\bm{y}|^2 d\gamma\right)^{1/2},
\end{gather}
where $\Pi(\mu,\nu)$ means all joint distributions with marginal distributions, $\mu$ and $\nu$ respectively.
Let $\tau$ be the timestep of the velocity-Verlet algorithm for the underdamped Langevin dynamics.
The following result indicates that our method is valid for capturing the finite time dynamics (we take the Langevin thermostat for illustration, discussions about other thermostats refer to \cite{JinLiXuZhao2020}).

\begin{theorem}\label{theorem1}
Let $(\bm{X}_i,\bm{V}_i)$ be the solution to the following second-order system,
	\begin{equation}\label{SDE1}
		\begin{aligned}
			&d\bm{X}_i=\bm{V}_idt,\\
			&m_id\bm{V}_i=(\bm{F}_i-\gamma \bm{V}_i)dt+\sqrt{2D}d\bm{W}_i,
		\end{aligned}
	\end{equation}
where $D=\gamma k_\text{B} T$, $\{\bm{W}_i\}_{i=1}^N$ are i.i.d. Wiener processes, and $\bm{X}_i$ and $\bm{V}_i$ denote the coordinates and the velocities of the $i$-th particle, respectively. The processes generated by the RBL are determined by the following SDEs,
 \begin{equation}\label{SDE2}
 	\begin{aligned}
 		&d\bm{\widetilde{X}}_i=\bm{\widetilde{V}}_idt,\\
 		&m_id\bm{\widetilde{V}}_i=(\bm{\widetilde{F}}_i-\gamma \bm{\widetilde{V}}_i)dt+\sqrt{2D}d\bm{W}_i.
 	\end{aligned}
 \end{equation}
Suppose that two SDEs Eq.\eqref{SDE1} and Eq.\eqref{SDE2} share the same initial data, and let $\bm{R}$ be the initial configuration of the system.
Suppose that the masses $m_i$ are bounded and the forces $\bm{F}_i$ are bounded and Lipschitz and $\mathbb{E}(\bm{\chi}_{i})=0$.
Then for any time $t^{*}>0$, there exists a constant $C(t^{*})$ independent of $N$ that
 \begin{equation}
 	\sup\limits_{\bm{R}} W_2(Q(\bm{R},\cdot),\widetilde{Q}(\bm{R},\cdot))\leq C(t^{*})\sqrt{\xi\tau+(1+D^2)\tau^2]}.
 \end{equation}
where $\xi=\left\|\mathbb{E}\left(\bm{\chi}_i^2\right)\right\|_{\infty}$,
$Q(\bm{R},\cdot)$ and $\widetilde{Q}(\bm{R},\cdot)$ are the transition probabilities of the SDEs of the direct truncation method Eq.\eqref{SDE1} and the RBL Eq.\eqref{SDE2}.
\end{theorem}
The proof of Theorem \ref{theorem1} is provided in Appendix \ref{app:Wassdist}. The molecular dynamics procedure using the RBL and
a heat bath with underdamped Langevin dynamics \cite{leimkuhler2013robust,duncan2017using,kopec2015weak} is summarized in Algorithm \ref{algo:RBL} below. 
\begin{algorithm}[H]
	\caption{(RBL)}\label{algo:RBL}
	\begin{algorithmic}[1]
		\STATE Input initial data: $N$, $V$, $T$, initialize particle positions and velocities, set $N_{t^{*}}$ be the total simulation steps, and choose batch size $p$
		\FOR {$n$ in $1:N_{t^{*}}$ }
		\STATE \quad Create the cell lists
 		\STATE \quad For each particle $i$, randomly choose $p$ particles in $\mathbb{I}$ if $N_{\mathbb{I}}>p$, otherwise choose all particles in $\mathbb{I}$
		\STATE \quad Calculate forces $\bm{F}_{i,\mathrm{c}}$ in the core region \eqref{coreF} and $\widetilde{\bm{F}}_{i,\mathrm{s}}$ in the shell zone \eqref{shellF}, respectively
		\STATE \quad Calculate $\widetilde{\bm{F}}_i=\bm{F}_{i,\mathrm{c}}+\widetilde{\bm{F}}_{i,\mathrm{s}}$ and subtract the average net force
		\STATE \quad Integrate Newton's equations with suitable integrate scheme and thermostat
		\ENDFOR
	\end{algorithmic}
\end{algorithm}

We now analyze the storage consumption and computational complexity of the RBL, in comparison with the classical direct truncation (DT) method combining the cell list with Verlet list.
Without loss of generality, we assume that particles have a uniform distribution and the mean particle density is $\rho$.
Both the RBL and the classical method need to allocate total $O(N)$ memories for storing the information of particles.
Regarding CPU memory usage for storing the neighbor list, $O(4\pi r_\mathrm{s}^3 \rho N/3)$ is required for the classical method whereas $O(4\pi r_\mathrm{c}^3 \rho N/3)$ for the RBL.
In practical simulations, the average number of particles within the cutoff radius for the classical method
is about several hundred, the RBL thus saves $(r_\mathrm{s}/r_\mathrm{c})^3$ memories compared to the classical method.
The calculation complexity per particle in classical method is $O(4\pi r_\mathrm{s}^3\rho/3)$, whereas the RBL reduced this cost to $O(4\pi r_\mathrm{c}^3\rho/3+p)$.  In the comparison below, if we safely adopt $r_\mathrm{s}/r_\mathrm{c}=2.5$ and appropriate $p$, both the storage saving and the speedup have about an order of magnitude improvement via the RBL.

\begin{remark}
	In this paper, we mainly focus on developing the theoretical and experimental knowledge of the RBL. The high-performance and parallel implementation of the RBL are left for future exploration.
	We note that the RBL will be friendly for parallelization. The strategies of MPI communication reported in Ref. \cite{plimpton1995fast}, including the domain decomposition, the load balance, and the diagnose, etc., can be naturally extended to the RBL. The vectorization of the RBL can refer to the algorithm in Ref.~\cite{pall2013flexible} which takes the benefit from the modern SIMD instruction set.
\end{remark}

\begin{remark}
	The LJ $12-6$ potential is arguably the most widely used pair potential in MD, whereas, as is pointed in \cite{wang2020lennard,hafskjold2019thermodynamic},
alternative short-range kernels are sometimes needed. The RBL can be naturally extended to these kernels, and the convergence proof provided in Appendix \ref{app:Wassdist} is available for general kernels.
\end{remark}

\begin{remark}
The RBL will bring in extra variance in the force term, leading to the numerical heating effect. Due to this reason, the RBL is not suitable for long time simulation under the NVE ensemble (similar to the work reported in \cite{JinLiXuZhao2020}) if without an appropriate symplectic scheme for time integration, but it should be good for NVT ensemble with thermostats.
\end{remark}

\section{Results}\label{sec:result}

In this section, we consider the LJ fluid under different conditions to validate the accuracy and efficiency of the proposed RBL.
We first calculate the equation of state of the LJ fluid by using the DT and the RBL with different parameters.
We then calculate physical properties of a multiphase coexistence state and compare the time consumption.
All calculations are implemented by using C++ and performed in a Linux system with Intel Xeon Scalable Cascade Lake 6248@2.5GHz, 1 CPU core and 4 GB memory.

We consider the standard $12-6$ type LJ potential Eq.\eqref{12-6} with different temperatures, densities and batch sizes in the RBL.
All quantities are provided in reduced units. We fix the particle number as $N=2000$.
The diameter of each particle is $\sigma=1$. The MD simulation utilizes a time step $\tau=0.002t_0$, where $t_0=\sigma\sqrt{m_0/k_B T}$
is the unit of time with the particle mass $m_0=1$ (LJ unit).
The simulation proceeds with velocity-Verlet scheme in the canonical ensemble using underdamped Langevin dynamics, where the thermal bath parameter takes $\gamma=1.0$.
In each simulation, we perform $0.2$ million timesteps for the equilibrium phase and $0.1$ million timesteps for the statistics to compute ensemble averages.
All contrastive curves are produced by the DT or the RBL.
The potential energy per particle and the system pressure at steady state with different temperatures and different number of $p$ are used to examine the accuracy.
Let the particle density be $\rho=N/L^3$ and the system be with periodic boundary condition.
With tail corrections, the potential energy and pressure formula are given by
\begin{equation}
U=\sum\limits_{r_{ij}<r_\mathrm{s}}4\left(\frac{\sigma^{12}}{r_{ij}^{12}}-\frac{\sigma^6}{r_{ij}^6}\right)+\dfrac{8}{3}\pi\rho\left(\dfrac{\sigma^{9}}{3r_\mathrm{s}^9}-\dfrac{\sigma^3}{r_\mathrm{s}^3}\right)
\end{equation}
and
\begin{equation}
P=\dfrac{\rho}{\beta}+\dfrac{8}{V}\sum\limits_{r_{ij}<r_\mathrm{s}} \left(\frac{2\sigma^{12}}{r_{ij}^{12}}-\frac{\sigma^{6}}{r_{ij}^6}\right)+\dfrac{16}{3}\pi\rho^2\left(\dfrac{2\sigma^{9}}{3r_\mathrm{s}^9}-\dfrac{\sigma^{3}}{r_\mathrm{s}^3}\right),
\end{equation}
where the rightmost terms of the two formulas are the tail corrections for $r_{ij}\geq r_\mathrm{s}$ \cite{frenkel2002book}, calculated by using the mean field theory, and are calculated with $r_\mathrm{s}=6$. Note that both $E$ and $P$ are computed after the end of the production step of the MD, based on the data collected.

\begin{figure*}[ht]	
	\centering
	\includegraphics[width=0.48\textwidth]{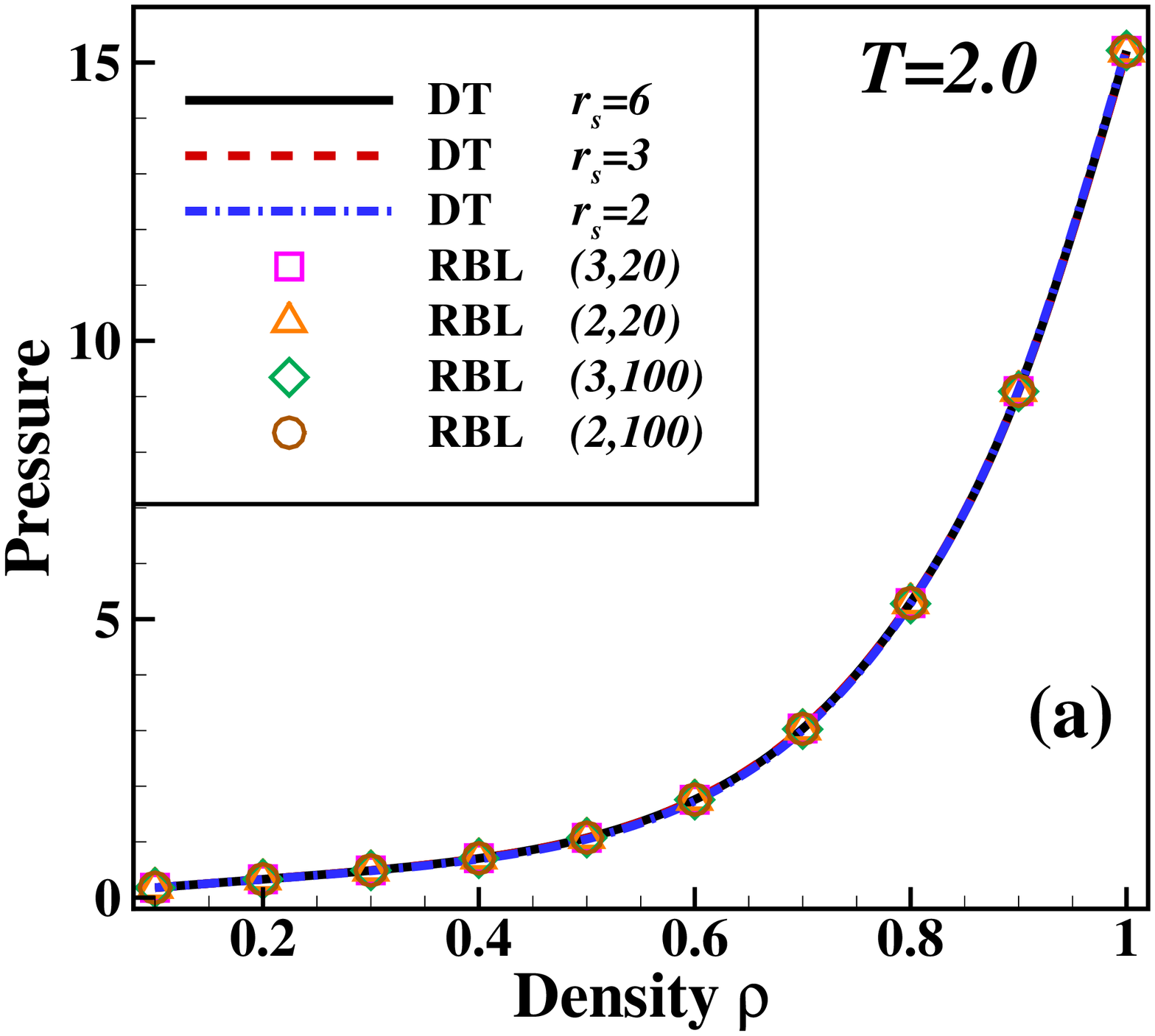}
	\includegraphics[width=0.48\textwidth]{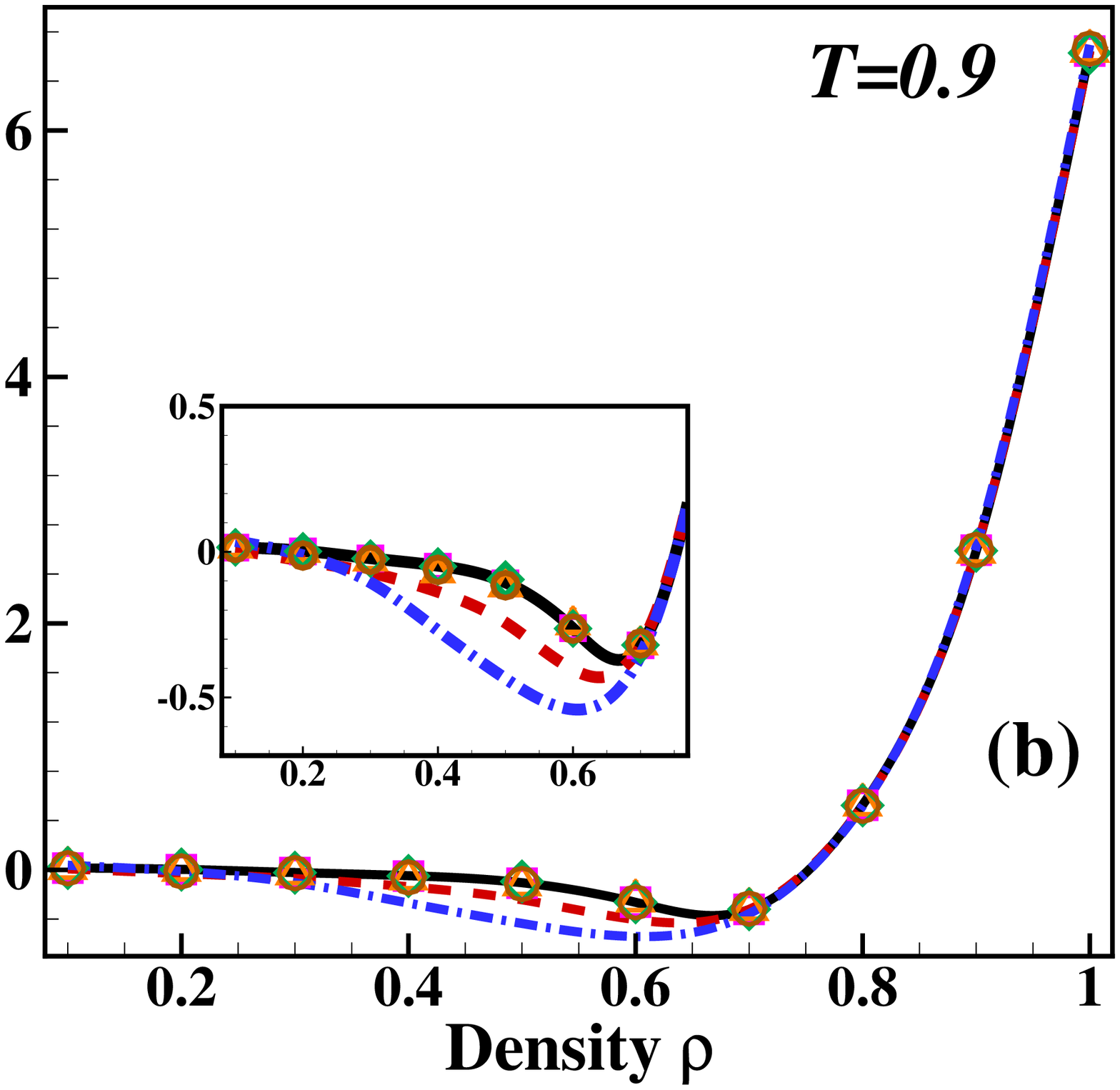}
	\includegraphics[width=0.48\textwidth]{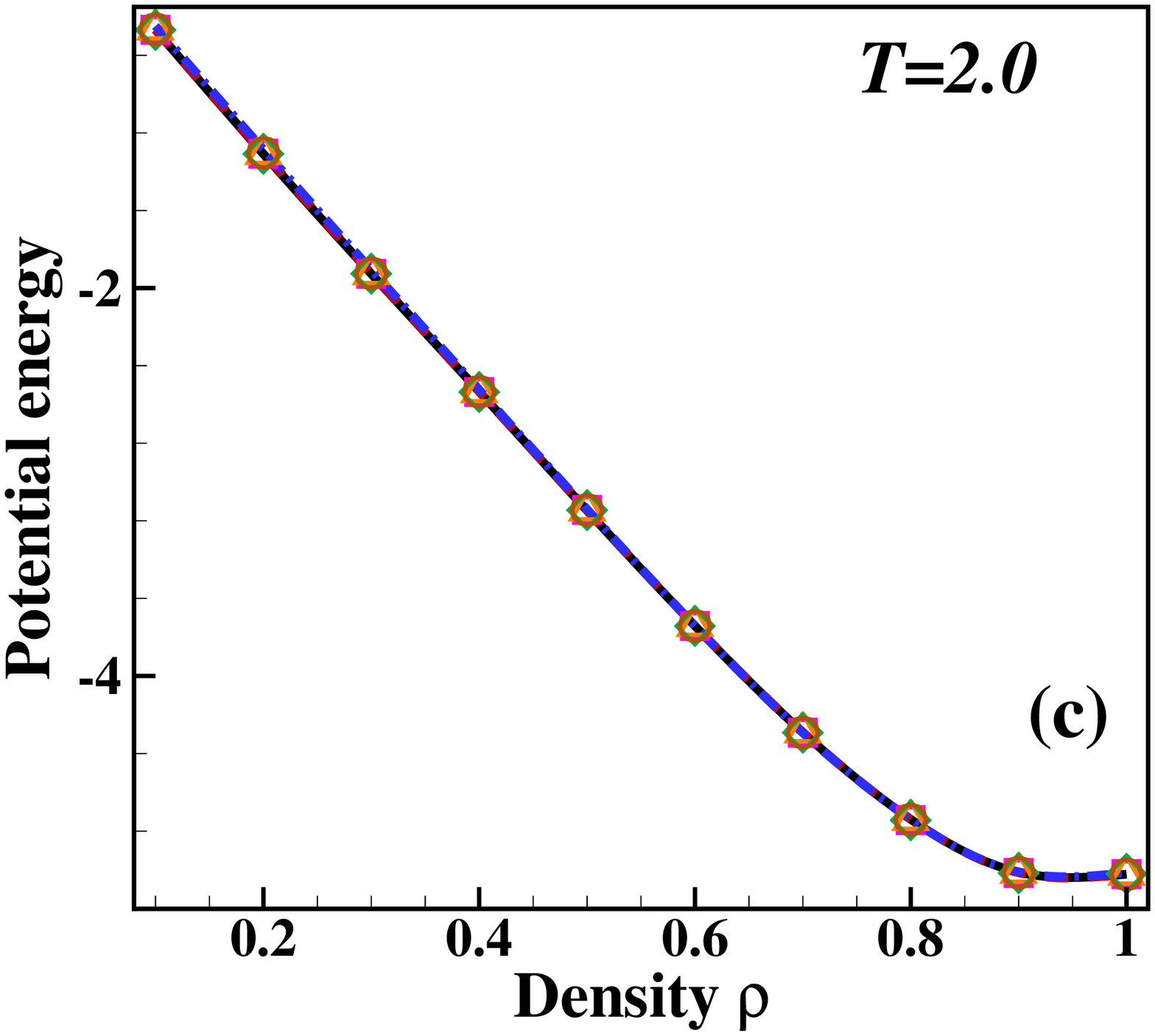}
	\includegraphics[width=0.48\textwidth]{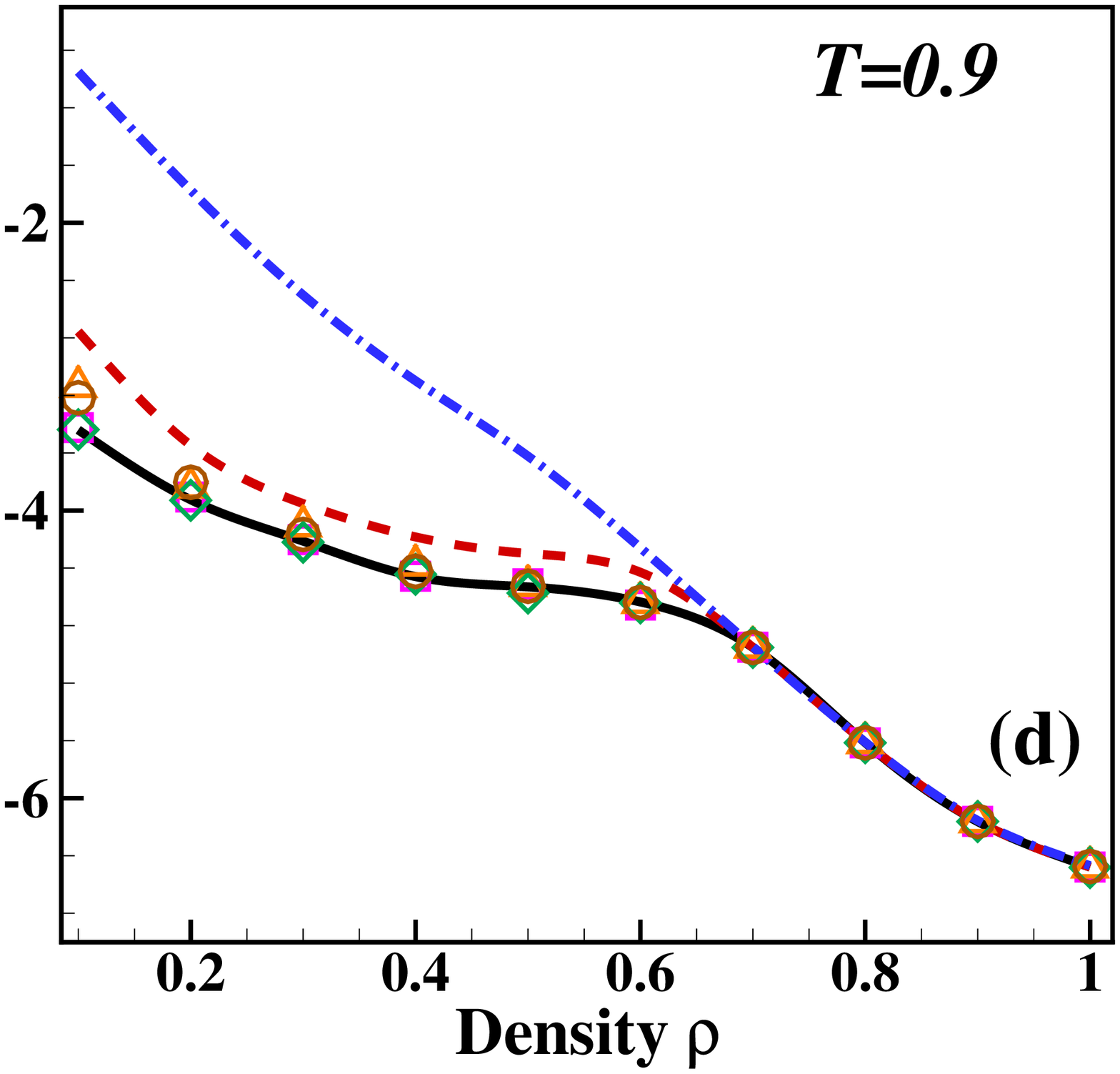}
	\caption{The ensemble average of pressure and potential energy per particle by simulation using the DT and the RBL with different parameters $(r_\mathrm{c},p)$ at temperature (a,c) $T=2.0$, and (b,d) $T=0.9$.}
	\label{fig:eqnstate}
\end{figure*}

For the DT, we use three cutoff radii, $r_\mathrm{s}=2,3$ and $6$, respectively.  And for the RBL, we take $r_\mathrm{s}=6$,
$r_\mathrm{c}=2$ and $3$, and $p=20$ and $100$.
The results are shown in Fig. \ref{fig:eqnstate}. We observe that both the DT and the RBL are accurate for systems at high temperature and high density.
But at low temperature and low particle density, the LJ fluid system can be at multiphase coexistence state.
In this situation, if one uses the DT, the stationary distribution can be very sensitive to the cutoff radius $r_\mathrm{s}$,
and the the stationary state keeps unchanged until $r_\mathrm{s}$ is large enough to provide accurate prediction.
From Fig. \ref{fig:eqnstate}, it is shown that the error produced by the RBL with different parameters is acceptable and it forms the gas-liquid state when $T=0.9$ and $\rho\leq 0.6$,
in agreement with the $r_\mathrm{s}=6$ case of the DT.
At low temperature and low density $\rho\leq 0.6$, the $r_\mathrm{s}=3$ case of the DT cannot converge to the true values of the potential energy and pressure.
The RBL with $r_\mathrm{c}=2$ and $p=20$ is already very accurate for both the energy and pressure calculations.
The errors with $r_\mathrm{c}=3$ are slightly smaller than those with $r_\mathrm{c}=2$ in the RBL, because the covariance when calculating $\widetilde{\bm{F}}_{i,\mathrm{s}}$ decreases quickly as $r_\mathrm{c}$ increases.

\begin{figure*}[ht]	
	\centering
	\includegraphics[width=0.48\textwidth]{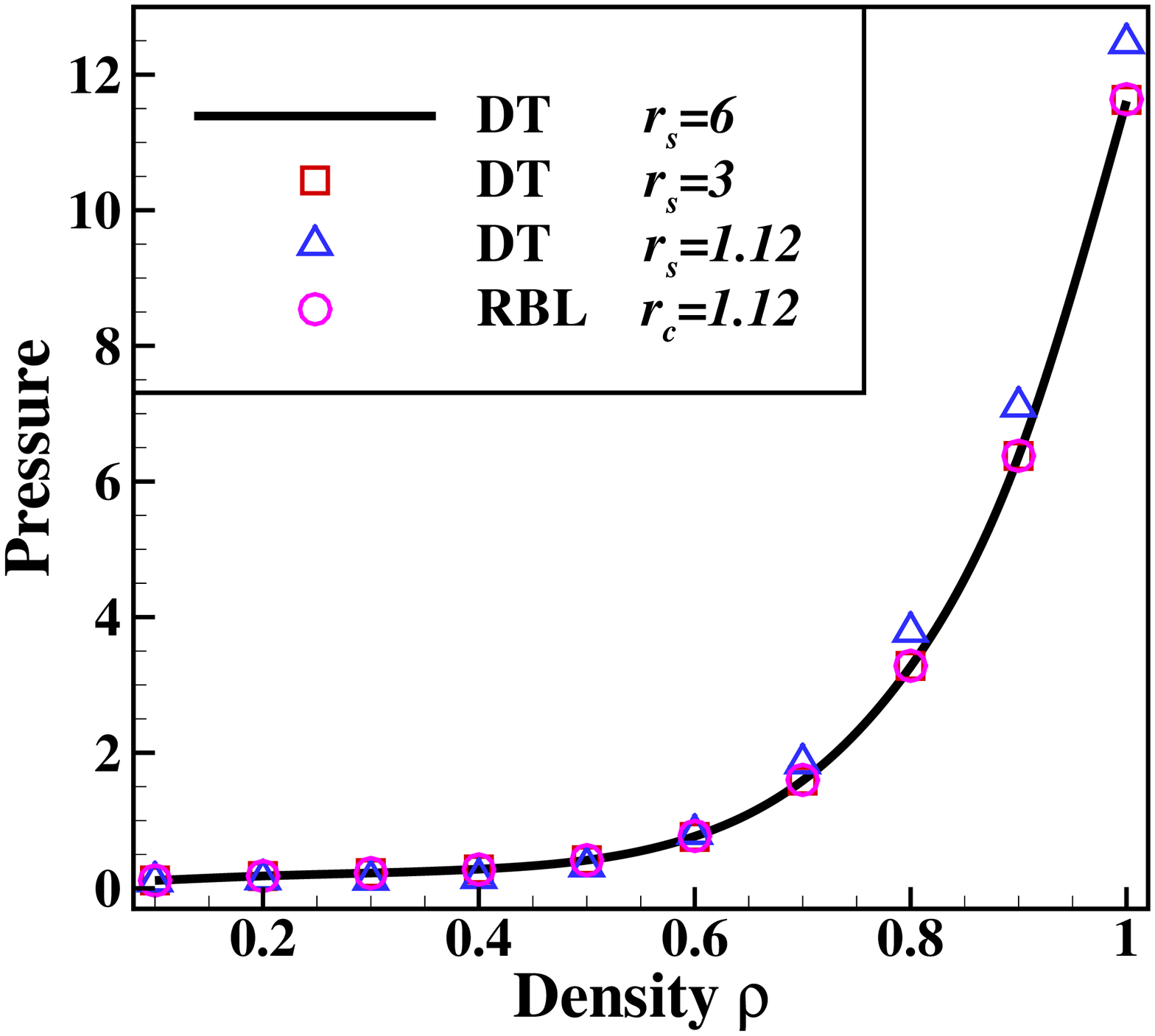}
	\includegraphics[width=0.48\textwidth]{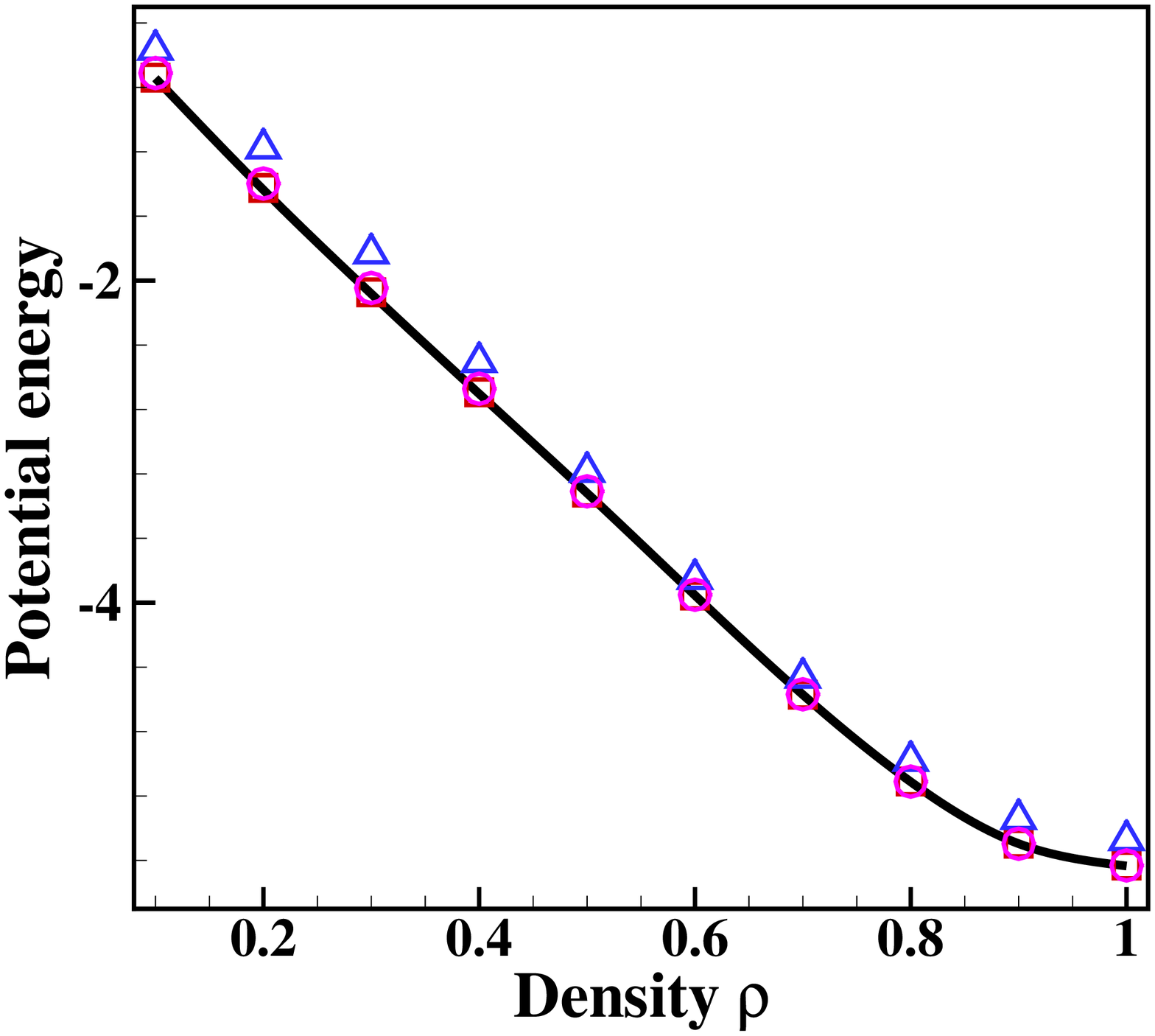}
	\caption{The ensemble average of pressure and potential energy per particle by simulation using the DT with different cutoff radius $r_{\mathrm{s}}$ and the RBL (with $r_\mathrm{s}=3$, $r_{\mathrm{c}}=2^{1/6}$ and $p=10$) in the homogeneous LJ fluid at $T=1.5$.}
	\label{fig:eqnstate3}
\end{figure*}

In many applications, the DT with cutoff radius $r_\mathrm{s}=3$ can provide enough accuracy, especially when the system is homogeneous. In such cases, the core radius of the RBL can be very small. To illustrate how small $r_\mathrm{c}$ can be, we simulate the LJ fluid system at $T=1.5$, and the results are shown in Fig. \ref{fig:eqnstate3}. Here, we choose $r_\mathrm{s}=6$ for the DT as the reference solution,
and we can see that the DT with $r_\mathrm{s}=3$ has already reproduced the solution. For the RBL, we take $r_\mathrm{c}=2^{1/6}$ ($\sim 1.12$ which is the radius of the LJ repulsive region), and the batch size $p=10$ in the shell zone of radius $3$. The DT result with $r_\mathrm{s}=2^{1/6}$ is also plotted, for which the potential energy and the pressure significantly deviate from the reference solutions,  with the maximum relative error about $20\%$.  The RBL with $r_\mathrm{s}=3$ and small core radius $r_{\mathrm{c}}=2^{1/6}$ is able to provide accurate results. 

\begin{figure*}[ht]	
	\centering
	\includegraphics[width=0.48\textwidth]{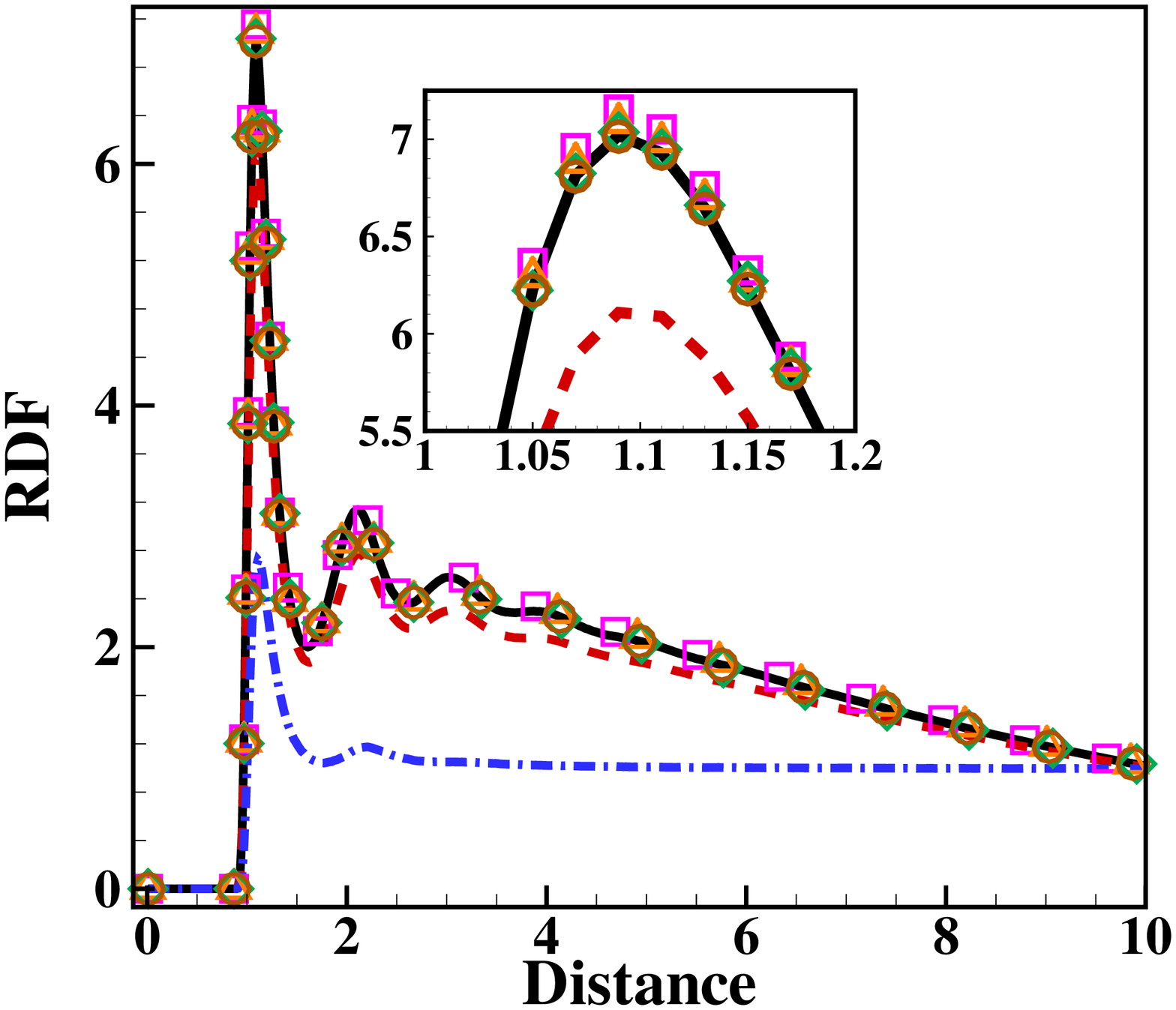}
	\includegraphics[width=0.48\textwidth]{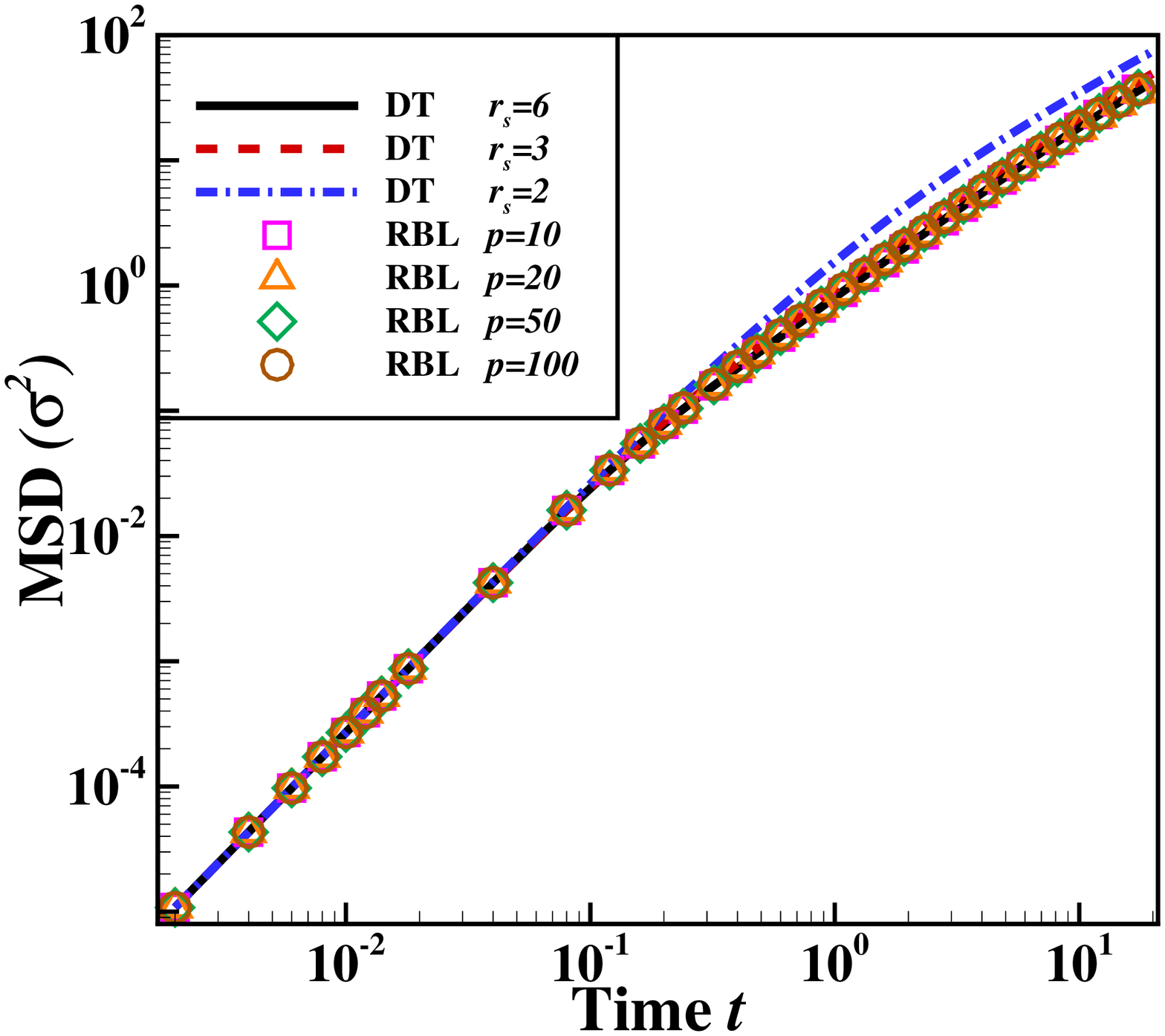}
	\caption{RDF and MSD predicted by simulations using the DT and the RBL ($r_\mathrm{s}=6$ and $r_\mathrm{c}=2$) with different parameters for $N=2000$ and $\rho=0.2$.}
	\label{fig:rdf}
\end{figure*}

Next, we calculate the radial distribution function (RDF) and the mean-squared displacement (MSD) to study the convergence of the batch size $p$ against by the result of the DT
for the case of gas-liquid coexistence. The RDF $g(r)$ at distance $r$ is defined by
\begin{equation}
	g(r)=\dfrac{1}{N\rho}\sum\limits_{i=1}^N\sum\limits_{j\neq i} \dfrac{\langle \delta(r_{ij}-r) \rangle}{4\pi r^2},
\end{equation}
where the bracket represents the ensemble average, and the MSD $\eta(t)$ at time $t$ reads
\begin{equation}
	\eta(t)=\langle |\bm{r}(t+t_0)-\bm{r}(t_0)|^2\rangle,
\end{equation}
with the bracket representing the ensemble average over $t_0$.
We fix $N=2000$, $\rho=0.2$ and $T=0.9$, while the system is at gas-liquid coexistence state, and consider $p=10,20,50$ and $100$, respectively in the RBL.
The simulation results produced by the RBL and DT are reported in Fig. \ref{fig:rdf}.
Using the DT, the system is still in gas phase if $r_\mathrm{s}=2$, while it becomes the coexistence of gas and liquid if $r_\mathrm{s}=3$ and $6$. The value of the first peak of RDF is $6.1$ when $r_\mathrm{s}=3$ while the referred true value is about $7.02$ as $r_\mathrm{s}=6$. The relatively accurate result could be obtained using the RBL even when $r_\mathrm{c}=2$ with a small batch size $p=10$.
In addition, it shows that the MSD $\eta(t)$ is quadratic of $t$ when $t$ is small and becomes linear when $t$ is large,
and the error in the DT with $r_\mathrm{s}=2$ is very clear.
Table \ref{tabl:pconverg} provides the values of the first peak of the RDF and the slope of the MSD depicted in Fig. \ref{fig:rdf},
which are calculated from the average of ten simulation runs. These data display the convergence of the RBL with the increasing $p$.

\begin{table}[ht]
	\caption{The value of the first peak of RDF and the slope of MSD, $\lim\limits_{t\rightarrow\infty} \eta(t)/t$ }
	\centering
	\begin{tabular}{c|ccc|cccc}
		\hline
		\multirow{2}{*}{}&
		\multicolumn{3}{c|}{DT}&\multicolumn{4}{c}{RBL ($r_\mathrm{s}=6$, $r_\mathrm{c}=2$)}\\\hline
		\cmidrule(lr){2-4} \cmidrule(lr){5-8}
		&$r_\mathrm{s}=6$&$r_\mathrm{s}=3$&$r_\mathrm{s}=2$&$p=10$&$p=20$&$p=50$&$p=100$\\\hline
		\midrule
		Peak  &7.0196&6.1095&2.6592&7.1532&7.0849&7.0366&7.0141\\\hline
		Slope  &2.1117&2.4668&3.7144&2.2367&2.0303&2.1669&2.1348\\\hline
	\end{tabular}\label{tabl:pconverg}
\end{table}

\begin{figure*}[ht]	
	\centering
	\includegraphics[width=0.48\textwidth]{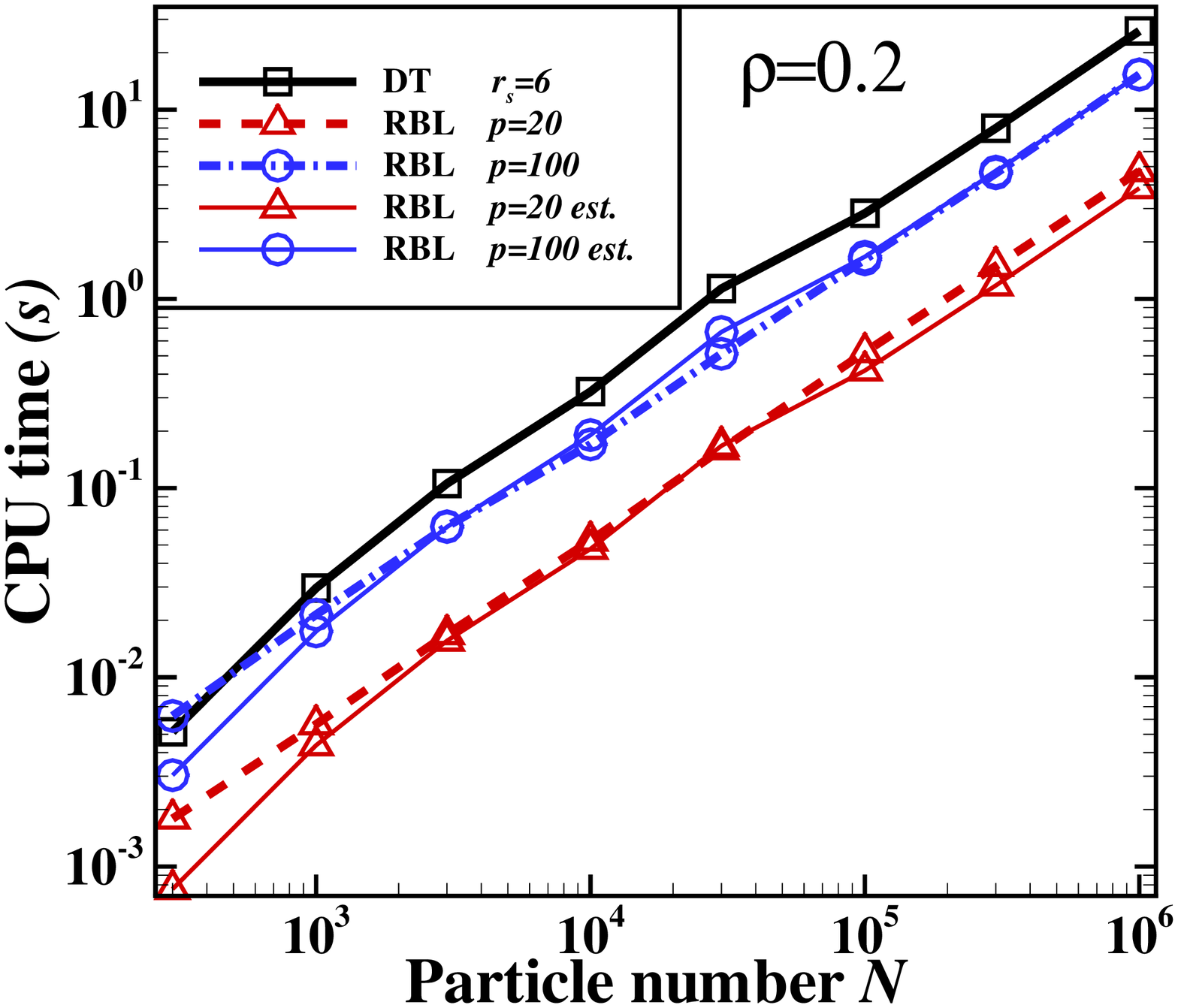}
	\includegraphics[width=0.48\textwidth]{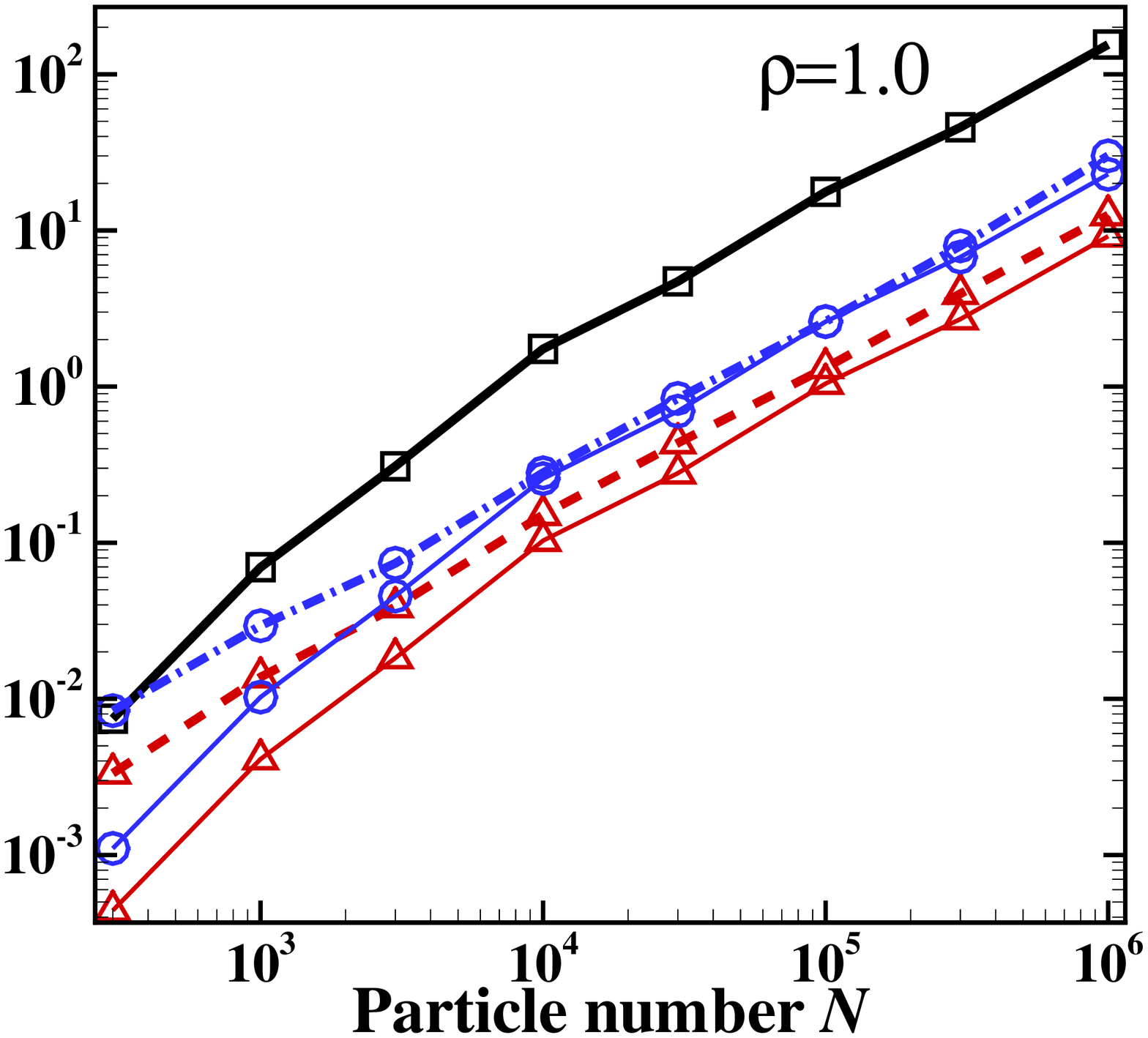}
	\caption{CPU time per step for the DT and the RBL ($r_\mathrm{s}=6$, $r_\mathrm{c}=2$) with increasing $N$ of two particle densities $\rho=0.2$ and $1.0$. The red/blue solid lines show the estimated CPU time of the RBL obtained from the DT by using the RBL/DT ratio from the theoretical  complexity.} 
	\label{fig:timecomp}
\end{figure*}

Then, we compare the CPU time of the DT and RBL. We set $r_\mathrm{s}=6$ for both methods and $r_\mathrm{c}=2$, $p=20$ and $100$ for the RBL.
Fig. \ref{fig:timecomp} illustrates the average CPU time per step of these two methods for different particle densities
($\rho=0.2$ for the left panel and $1.0$ for the right panel) as a function of the number of particles (up to $N = 10^6$).
We observe linear growth of the CPU time with $N$ for large particle number for all cases. The estimated CPU times of the RBL with different $p$ are calculated by using the DT time and the theoretical RBL/DT ratio of the complexity discussed in Section \ref{Sec::Med},
and our results show the correctness of the complexity analysis.
Under the low-density case with $\rho=0.2$, the computational time of the RBL with $p=20$ has about $6-7$ times faster than that of the DT method.
This speedup moves up to over 10 times when the density of the system becomes $\rho=1.0$, indicating the efficiency of the RBL for systems with
high density is even more significant.

\begin{figure*}[htbp]	
	\centering
	\includegraphics[width=0.48\textwidth]{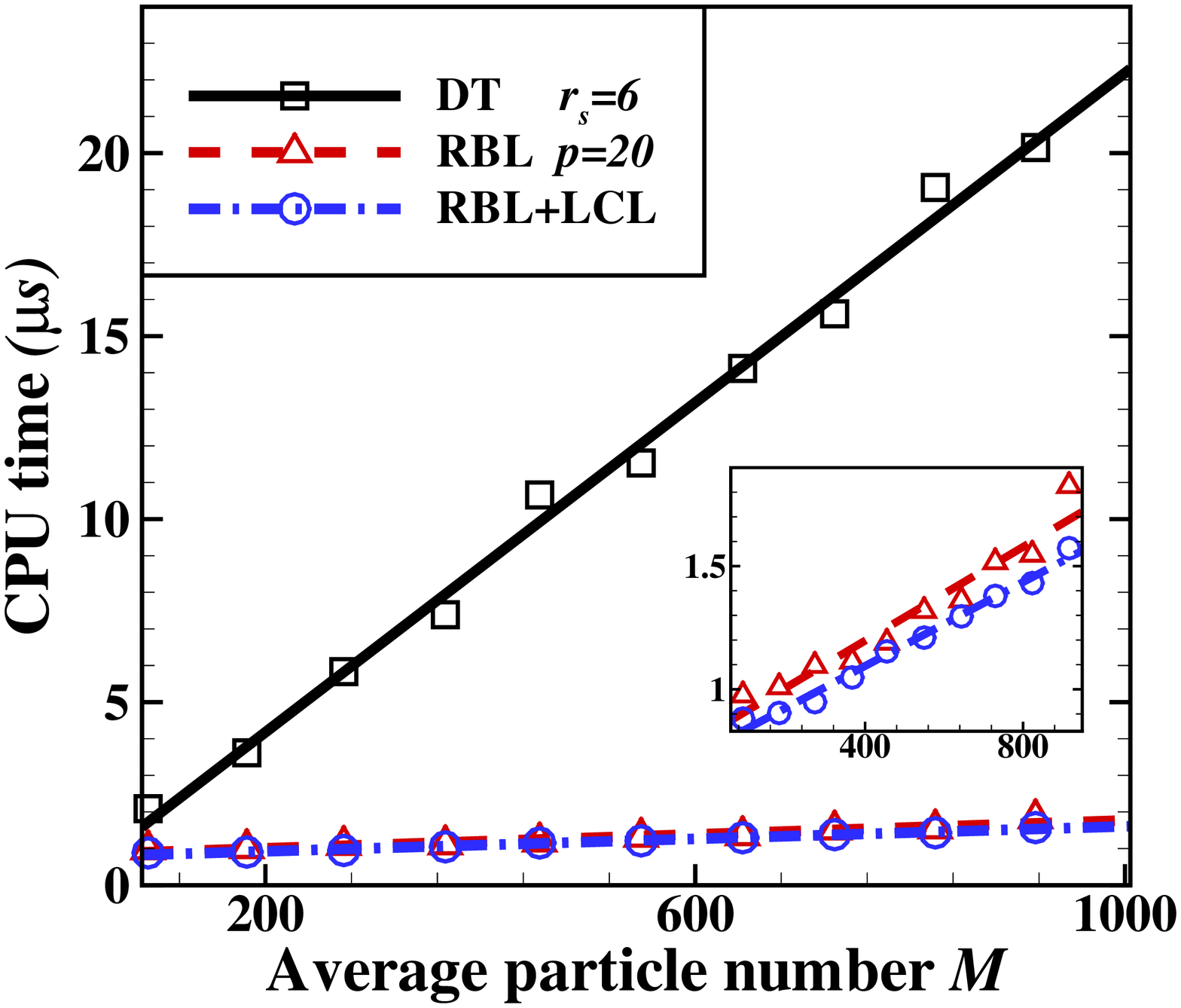}
	\caption{CPU time performance as function of the average number of interacting particles using the DT, the RBL, and the RBL combining with the LCL.}
	\label{fig:timecomp2}
\end{figure*}

Finally, we examine the comparison on time performance between the DT, the RBL, and the RBL accelerated by the classical LCL \cite{welling2011efficiency}.
To assess the efficiency, we perform a set of simulations with varying average number of interactions per particle $M$ which is a value depending on the
particle density. Here $M$ is calculated by the DT with $r_\mathrm{s}=6$. The CPU time is measured by the average time per step and particle in the unit of
microseconds, i.e., the total simulation time divided by the product of simulation steps and particle number.
We set $T=2.0$, $N=10000$, $r_\mathrm{s}=6$ for all the cases, and $r_\mathrm{c}=2$ and $p=20$ for the RBLs. The results are presented in Fig.~\ref{fig:timecomp2},
showing that the RBL is very efficient and greatly accelerates the calculations, especially for systems with high particle density.
The LCL technique can further provide about $15\%$ acceleration. We remark that the efficiency of the algorithms depends on actual implementations of these methods and in this case the implementations were carried out by different programmers. In particular, our implementations of the RBL and the LCL method are not fully optimized, and we believe that better performance should be achieved by using state-of-the-art LCL techniques and we will continuously improve it in our future work. 

\section{Conclusion}

In summary, we have developed a novel neighbor-list algorithm for MD with short-range interactions. The RBL method benefits from a random mini-batch idea for calculating the force in the shell zone.
The accuracy and efficiency of the proposed algorithm are examined by reproducing the physical properties of the LJ fluid, particularly the heterogeneous system in the case of gas-liquid coexistence.
The RBL can be applied into other short-range interactions such as the error complementary function over $r$, $\text{erfc}(\alpha r)/r$, which is the short-range component of Coulomb interactions.
Thus, the RBL can be used to treat the short-range part of the random batch Ewald method for Coulomb interactions \cite{JinLiXuZhao2020}, and with these treatment the MD simulation
is expected to simulate larger systems. The research of all-atom simulations coupling the random batch Ewald and RBL shall be studied in our subsequent works. Moreover, if the system is partially periodic in some directions with dielectric interfaces (e.g., the solid/liquid/vapor interface) \cite{liang2020harmonic,tyagi2008electrostatic,hu2014symmetry}, the extension of our algorithm is also straightforward.  

\section*{Acknowledgements}
The work was supported by  NSFC (grant No. 12071288), Shanghai Science and Technology Commission  (grant No. 20JC1414100) and the HPC center of Shanghai Jiao Tong University.

\section*{ Data Availability}
The data that support the findings of this study are available from the corresponding author upon reasonable request.

\appendix

\section*{Appendix}

\section{Expectation and variance of the forces}\label{app:forcevar}

The force on particle $i$ with the cutoff radius $r_\mathrm{s}$ can be written as the summation of the core and shell
contributions,
$\bm{F}_{i}=\bm{F}_{i,\mathrm{c}}+\bm{F}_{i,\mathrm{s}}.$
In the RBL, the force on particle $i$ is
$\widetilde{\bm{F}}_i=\bm{F}_{i,\mathrm{c}}+\widetilde{\bm{F}}_{i,\mathrm{s}}-\bm{F}_{\text{cor}}$,
where the force due to the shell particles $\widetilde{\bm{F}}_{i,\mathrm{s}}$ is defined in Eq. \eqref{shellF} and $\bm{F}_{\text{cor}}$ is the average net force,
\begin{equation}
\bm{F}_{\text{cor}}=\dfrac{1}{N}\sum\limits_{i=1}^N (\bm{F}_{i,\mathrm{c}}+\widetilde{\bm{F}}_{i,\mathrm{s}}).
\end{equation}
Let $I_{ij}$ be the indicative function which is one if $j\in B(i)$ and zero otherwise. It is easy to obtain,
\begin{equation}
	\mathbb{E}(\widetilde{\bm{F}}_{i,\mathrm{s}})=\frac{N_{\mathbb{I}}}{p}\sum\limits_{r_\mathrm{c}<r_{ij}<r_\mathrm{s}} \bm{f}_{ij}\mathbb{E}(I_{ij})=\bm{F}_{i,\mathrm{s}}.
\end{equation}
The average net force has zero expectation due to the force balance $\bm{f}_{ij}=-\bm{f}_{ji}$.
And therefore, the expectation of the force error is
\begin{equation}
	\mathbb{E}(\bm{\chi}_{i})=\mathbb{E}(\bm{F}_{i,\mathrm{s}}-\widetilde{\bm{F}}_{i,\mathrm{s}}+\bm{F}_{\text{cor}})=0,
\end{equation}
i.e., the force by the RBL is unbiased.

The variance of the force approximation can be written as
\begin{equation}\label{variance}
\var{\bm{\chi}_i}=\mathbb{E}(\bm{\widetilde{F}}_{i,\mathrm{s}}^2-\bm{F}_{i,\mathrm{s}}^2+\bm{F}_{\text{cor}}^2-2\bm{\widetilde{F}}_{i,\mathrm{s}}\bm{F}_{\text{cor}}).
\end{equation}
We discuss the four contributions separately. One has
\begin{equation}\label{corr}
	\begin{split}
		\mathbb{E}(\bm{\widetilde{F}}_{i,\mathrm{s}}\bm{F}_{\text{cor}})=&\dfrac{1}{N}\mathbb{E}\left[\bm{\widetilde{F}}_{i,\mathrm{s}}\sum\limits_{k=1}^N (\bm{F}_{k,\mathrm{c}}+\widetilde{\bm{F}}_{k,\mathrm{s}})\right]\\
		=&\dfrac{1}{N}\mathbb{E}(\bm{\widetilde{F}}_{i,\mathrm{s}})\mathbb{E}\left[\sum\limits_{k\neq i}^N (\bm{F}_{k,\mathrm{c}}+\widetilde{\bm{F}}_{k,\mathrm{s}})+\bm{F}_{i,\mathrm{c}}\right]\\
		&+\dfrac{1}{N}\mathbb{E}(\widetilde{\bm{F}}_{i,\mathrm{s}}^2)\\
		=&\dfrac{1}{N}\left[\mathbb{E}\left(\widetilde{\bm{F}}_{i,\mathrm{s}}^2\right)-\bm{F}_{i,\mathrm{s}}^2\right],
	\end{split}
\end{equation}
where the second identity employs the fact of $\bm{f}_{ij}=-\bm{f}_{ji}$ and the mutual independence of forces on particles.
The variance of $\bm{F}_{\text{cor}}$ reads	
\begin{equation}\label{square}
	\begin{split}
		\mathbb{E}(\bm{F}_{\text{cor}}^2)=&\dfrac{1}{N^2}\mathbb{E}\left[\sum_{i,j=1}^{N}(\bm{F}_{i,\mathrm{c}}+\widetilde{\bm{F}}_{i,\mathrm{s}})(\bm{F}_{j,\mathrm{c}}+\widetilde{\bm{F}}_{j,\mathrm{s}})\right]\\
		=&\dfrac{1}{N^2}\sum\limits_{i=1}^{N}\left(\mathbb{E}(\widetilde{\bm{F}}_{i,\mathrm{s}}^2)-\bm{F}_{i,\mathrm{s}}^2\right).
	\end{split}
\end{equation}
The expectations of $\widetilde{\bm{F}}_{i,\mathrm{s}}^2$ and $\bm{F}_{i,\mathrm{s}}^2$ are
\begin{equation}
\begin{split}
\mathbb{E}(\widetilde{\bm{F}}_{i,\mathrm{s}}^2)=&\dfrac{N_{\mathbb{I}}^2}{p^2}\mathbb{E}\left[\left(\sum\limits_{i\neq j} \bm{f}_{ij}I_{ij}\right)^2\right]\\
=&\dfrac{N_{\mathbb{I}}^2}{p^2}\mathbb{E}\left[\sum\limits_{i\neq j}\bm{f}_{ij}^2I_{ij}\right]+\dfrac{N_{\mathbb{I}}^2}{p^2}\mathbb{E}\left[\sum\limits_{i\neq j} \sum\limits_{i\neq k}\bm{f}_{ij}\bm{f}_{ik}I_{ij}I_{ik}\right]\\
=&\dfrac{N_{\mathbb{I}}}{p}\left[\sum\limits_{i\neq j}\bm{f}_{ij}^2+\dfrac{p-1}{N_{\mathbb{I}}-1}\sum\limits_{i\neq j} \sum\limits_{i\neq k} \bm{f}_{ij}\bm{f}_{ik}\right],\\
\mathbb{E}(\bm{F}_{i,\mathrm{s}}^2)=&\sum\limits_{i\neq j}\bm{f}_{ij}^2+\sum\limits_{i\neq j} \sum\limits_{i\neq k} \bm{f}_{ij}\bm{f}_{ik},
\end{split}
\end{equation}
where the sum of force is over particles with $i\neq j\neq k$ and $r_\mathrm{c}\leq r_{ij},r_{ik} \leq r_\mathrm{s}$, thus we obtain
\begin{equation}\label{substraction}
	\mathbb{E}(\widetilde{\bm{F}}_{i,\mathrm{s}}^2-\bm{F}_{i,\mathrm{s}}^2)\lesssim\dfrac{N_{\mathbb{I}}-p}{p}\sum\limits_{\substack{r_\mathrm{c}\leq r_{ij} \leq r_\mathrm{s}\\i\neq j}}\bm{f}_{ij}^2.
\end{equation}

Finally, we substitute Eq.\eqref{corr}, \eqref{square} and \eqref{substraction} into Eq.\eqref{variance}, and obtain the following estimate:
\begin{equation}
\var(\bm{\chi}_{i})\lesssim \dfrac{N_{\mathbb{I}}-p}{N^2p}\left[\sum\limits_{i\neq j}C(N)\bm{f}_{ij}^2 +\sum\limits_k \sum\limits_{k\neq j}\bm{f}_{kj}^2\right],
\end{equation}
where $C(N)=N^2-2N$ and the second term is derived from the square expectation of correction term.

For the LJ fluid system in section III, we assume that the system is homogeneous with particle density $\rho$. An asymptotic bound of the leading term of $\var(\bm{\chi}_i)$ is given by
\begin{equation}\label{var}
\begin{aligned}
\var(\bm{\chi}_i)\lesssim&\dfrac{(N_{\mathbb{I}}-p)}{p}\left[\int_{r_\mathrm{c}}^{r_\mathrm{s}} 4\pi \rho r^2\left(\dfrac{24}{r^7}-\dfrac{48}{r^{13}}\right)^2 dr\right]\\ \lesssim& \dfrac{(N_{\mathbb{I}}-p)\rho}{p}r_\mathrm{c}^{-11}.
\end{aligned}
\end{equation}
This clearly shows that for pure LJ system the variance of $\bm{\chi}_{i}$ decays rapidly with the increase of $r_\mathrm{c}$.

\section{Proof of Theorem \ref{theorem1}}\label{app:Wassdist}

Let $(\bm{X},\bm{V})$ be the solution to the desired second order system Eq.\eqref{SDE1}, where $\bm{X}=\{\bm{X}_i\}_{i=1}^N$ and
$\bm{V}=\{\bm{V}_i\}_{i=1}^N$ are $3N$-dimensional column vectors.
Let $(\widetilde{\bm{X}},\widetilde{\bm{V}})$ be the solution to Eq.\eqref{SDE2} by the RBL.
Let us define $\bm{Y}=[\bm{X}'~(\bm{M}\bm{V})']'$ and $\widetilde{\bm{Y}}=[\widetilde{\bm{X}}'~(\bm{M}\widetilde{\bm{V}})']'$,
where $\bm{M}$ is $3N\times3N$ diagonal mass matrix and $'$ indicates the transpose.
Then we rewrite Eq.\eqref{SDE1} and Eq.\eqref{SDE2} into the following form, for $t\in[0,t^{*}]$,
\begin{equation}
	\begin{split}
	d\bm{Y}=(\bm{G}_1(\bm{Y})+\bm{G}_2(\bm{Y}))dt+\bm{G}_3 d\bm{W},\\
	d\widetilde{\bm{Y}}=(\bm{G}_1(\widetilde{\bm{Y}})+\widetilde{\bm{G}}_2(\widetilde{\bm{Y}}))dt +\bm{G}_3 d\bm{W},
	\end{split}
\end{equation}
where $$\bm{G}_1(\bm{Y})=\begin{bmatrix}\bm{0}&\bm{M}^{-1}\\\bm{0}&-\gamma \bm{M}^{-1}\end{bmatrix}\bm{Y}, ~~~\bm{G}_2(\bm{Y})=\begin{bmatrix}\bm{0}\\\bm{F}(\bm{Y}) \end{bmatrix},$$ $$\widetilde{\bm{G}}_2(\widetilde{\bm{Y}})=\begin{bmatrix}\bm{0}\\\widetilde{\bm{F}}(\widetilde{\bm{Y}})\end{bmatrix},~~~
\bm{G}_{3}=\begin{bmatrix}\bm{0}\\\sqrt{2D}\bm{I}\end{bmatrix}.$$
Here, $\bm{F}(\bm{Y})=\{\bm{F}_i\}_{i=1}^N$ and $\widetilde{\bm{F}}(\widetilde{\bm{Y}})=\{\widetilde{\bm{F}}_i\}_{i=1}^N$ are
force vectors on particles,
and $d\bm{W}$ is $3N$-dimensional standard Brownian motion.

We define $t_k=k\tau$ with $\tau$ the discrete step size and let $\bm{\delta Y}(t)=\bm{Y}(t)-\widetilde{\bm{Y}}(t)$. We consider
numerical results of Eq.\eqref{SDE1} and Eq.\eqref{SDE2} for $t\in[t_k,t_{k+1}]$ and apply the standard coupling technique
to estimate the error of the numerical solution $\widetilde{\bm{Y}}(t)$ compared with $\bm{Y}(t)$. By the It\^o's calculus, it is found that,
\begin{equation}\label{Ito1}
	\begin{aligned}
		\dfrac{d}{dt}\mathbb{E}|\bm{\delta Y}(t)|^2=&2\mathbb{E}[\bm{\delta Y}(t)]
		\cdot [\bm{G}_1(\bm{Y}(t))-\bm{G}_1(\widetilde{\bm{Y}}(t_k)) \\&+\bm{G}_2(\bm{Y}(t))-\widetilde{\bm{G}}_2(\widetilde{\bm{Y}}(t_k))].
	\end{aligned}
\end{equation}
Let $\bm{\delta G}_{1}(t)=\bm{G}_{1}\left(\bm{Y}(t)\right)-\bm{G}_{1}(\widetilde{\bm{Y}}(t_k))$ and
$\bm{\delta G}_2(t) =\bm{G}_2\left(\bm{Y}(t)\right)-\widetilde{\bm{G}}_2(\widetilde{\bm{Y}}(t_k))$. The It\^o's formula gives
\begin{equation}\label{Y1}
\bm{\delta Y}(t)=\bm{\delta Y}(t_k)+\int_{t_k}^{t} \left[\bm{\delta G}_1(s)+\bm{\delta G}_2(s)\right]ds
\end{equation}
and
\begin{equation}\label{G2}
\begin{aligned}
\bm{G}_{\ell}(\bm{Y}(t))=\bm{G}_{\ell}(\bm{Y}(t_k))&+\int_{t_k}^{t}d\bm{Y}\cdot\nabla \bm{G}_{\ell}(\bm{Y}(s))\\
&+\int_{t_k}^{t} D\Delta \bm{G}_{\ell}(\bm{Y}(s))ds,
\end{aligned}
\end{equation}
for $\ell=1,2$. Substituting Eq.\eqref{Y1} and \eqref{G2} into Eq.\eqref{Ito1}, one has
\begin{equation}\label{final1}
\begin{aligned}
\dfrac{d}{dt}\mathbb{E}|\bm{\delta Y}(t)|^2=&2\mathbb{E} [\bm{\delta Y}(t_k)\bm{\delta G}(t)]\\
&+2\mathbb{E}\left[\int_{t_k}^{t} \bm{\delta G}(s) ds\bm{\delta G}(t)\right],
\end{aligned}
\end{equation}
where $\bm{\delta G}(t)=\bm{\delta G}_1(t)+\bm{\delta G}_2(t)$.

Since the masses $m_i$ are bounded, the forces $\bm{F}_i$ are bounded and Lipschitz,
$\mathbb{E}(\bm{\chi}_i)=0$ and $\mathbb{E}(\bm{\chi}_i^2)$ are bounded for all $i$,
by Eq. \eqref{G2}, we obtain the following estimate with constant $C$,
\begin{equation}\label{et1}
	\mathbb{E} \left(\bm{\delta Y}(t_k)\bm{\delta G}_{\ell}(t)\right)\leq C\left[\left\|\bm{\delta Y}(t_k)\right\|^2 +||\bm{\delta Y}(t_k)||(1+D)\tau\right],
\end{equation}
where $\|\cdot\|:=\sqrt{\mathbb{E}|\cdot|^2}$ is the $L^2(\mathbb{P})$ norm. And
\begin{equation}\label{et2}
\begin{split}
&\mathbb{E} \left(\int_{t_k}^{t} \bm{\delta G}_{\ell_1}(s)ds\cdot \bm{\delta G}_{\ell_2}(t)\right)\leq C\left\|\bm{\delta Y}(t_k)\right\|^2\tau\\
&+C(||\bm{\delta Y}(t_k)||)(1+D)\tau^2+CD\tau^2,\\
\end{split}
\end{equation}
where $(\ell_1,\ell_2)$ could be $(1,1)$, $(1,2)$ and $(2,1)$, and
\begin{equation}\label{et3}
	\mathbb{E} \left(\int_{t_k}^{t}\bm{\delta G}_2(s)ds \cdot \bm{\delta G}_2(t)\right)\leq 2\max\left\{\mathbb{E}\left(\bm{\chi}_i^2\right)\right\}\tau+CD\tau^2.
\end{equation}

Finally, substituting Eqs. \eqref{et1}-\eqref{et3} into Eq.\eqref{final1} and recalling that $\xi=\left\|\mathbb{E}\left(\bm{\chi}_i^2\right)\right\|_{\infty}$, one has
\begin{equation}
\dfrac{d}{dt}\|\bm{\delta Y}(t)\|^2\leq C\left(\|\bm{\delta Y}(t_k)\|^2+(1+D^2)\tau^2\right)+2\xi\tau
\end{equation}
for all $t\in[t_k,t_{k+1}]$. Hence, we have
\begin{equation}
\begin{aligned}
\left\|\bm{\delta Y}(t_{k+1})\right\|^2\leq& (1+C\tau)\left\|\bm{\delta Y}(t_k)\right\|^2\\
&+2\xi\tau^2+C(D^2+1)\tau^3.
\end{aligned}
\end{equation}
By the Gronwall inequality, the convergence reads
\begin{equation}
\begin{aligned}
||\bm{\delta Y}(t^{*})||^2\lesssim C(t^{*})\left[\xi\tau+(1+D^2)\tau^2\right],
\end{aligned}
\end{equation}
where $C(t^{*})$ is a constant depending on $t^{*}$. Let $\bm{R}$ be the initial configuration of the system, the Wasserstein distance between the SDEs of the DT and the RBL shows
\begin{equation}\label{final}
\sup\limits_{\bm{R}} W_2(Q(\bm{R},\cdot),\widetilde{Q}(\bm{R},\cdot))\leq C(t^{*})\sqrt{\xi\tau+(1+D^2)\tau^2}.
\end{equation}

As analyzed in Appendix \ref{app:forcevar}, for the LJ fluid system, assume that the system is homogeneous with particle density $\rho$,
one has $\xi\lesssim ((N_{\mathbb{I}}-p)/p)\rho r_\mathrm{c}^{-11}.$ The $O(\tau^{1/2})$ term in Eq.\eqref{final} vanishes with the increase of $r_\mathrm{c}$,
thus $\sup_{\bm{R}} W_2(Q(\bm{R},\cdot),\widetilde{Q}(\bm{R},\cdot))\sim O(\tau)$.


\end{document}